\documentclass[pra,onecolumn,showpacs,eqsecnum,preprintnumbers,amsmath,amssymb,superscriptaddress]{revtex4}

\usepackage{bm}
\usepackage{graphicx}
\usepackage{amsbsy} 
\usepackage{amsmath}
\usepackage{amsfonts}
\usepackage{amsthm}

\newcommand{\vecb}{\bm}

\begin{document}

\newtheorem{theorem}{Theorem}
\newtheorem{lemma}[theorem]{Lemma}
\newtheorem{corollary}[theorem]{Corollary} 
\newtheorem{conjecture}[theorem]{Conjecture}
\newtheorem{claim}[theorem]{Claim}

\newtheorem{definition}[theorem]{Definition}

\newtheorem{remark}[theorem]{Remark}
\newtheorem{example}[theorem]{Example}

\title{Duality for monogamy of entanglement}

%\author{Somshubhro Bandyopadhyay\inst{1}, Gilad Gour\inst{2}\and Barry C.\ Sanders\inst{1,3}}
%\institute{Institute for Quantum Information Science, University of Calgary, Alberta T2N 1N4, Canada
%\and Department of Mathematics, University of California/San Diego, La Jolla, California 92093
%\and Centre of Excellence for Quantum Computer Technology, Macquarie University, 
%	Sydney, New South Wales 2109, Australia}

\author{Somshubhro Bandyopadhyay}\email{som@qis.ucalgary.ca}
\affiliation{Institute for Quantum Information Science, University of Calgary, Alberta T2N 1N4, Canada}
\author{Gilad Gour}\email{ggour@math.ucsd.edu}
\affiliation{Department of Mathematics, University of California/San Diego, 
       La Jolla, California 92093-0112}
\author{Barry C. Sanders}
\affiliation{Institute for Quantum Information Science, University of Calgary, Alberta T2N 1N4, Canada}
\affiliation{Centre of Excellence for Quantum Computer Technology, Macquarie University, 
	Sydney, New South Wales 2109, Australia}

\date{\today}

\begin{abstract}
We establish duality for monogamy of entanglement: whereas monogamy of entanglement
inequalities provide an upper bound for bipartite sharability of entanglement in a multipartite
system, we prove that the same quantity provides a \emph{lower} bound for 
distribution of bipartite entanglement in a multipartite system. 
Our theorem for monogamy of entanglement is used to establish relations between
bipartite entanglement that separate one qubit from the rest vs separating
two qubits from the rest.
\end{abstract}


\maketitle

\section{Introduction}

In contrast to classical multipartite systems, which can enjoy arbitrary correlations between
components, shared entanglement is restricted in a multipartite system. In its most restrictive form, a pair of components that are maximally entangled cannot share
entanglement nor classical correlations with any part of the rest of the system, hence
the term `monogamy'~\cite{Ter03,Cof00,Koa04,Osb05}. Monogamy of entanglement (MoE) is interesting
both fundamentally as a quintessential quantum property, and also these restrictions on
sharing real or effective entanglement in quantum key distribution guarantees security
of the classical shared random key~\cite{Ren05}.

Whereas MoE inequalities provide upper bounds for sharability
of entanglement in subsystems~\cite{Cof00,Osb05}, we prove that this bound 
also acts as a lower bound (conjectured in~\cite{Gou05})
for distribution of entanglement, or `entanglement of assistance'~\cite{DiV99,Coh98,Lau01},
to a target pair~A and~B.
This distribution of entanglement is performed by the rest of the subsystems,
who perform collective operations to assist~A and~B to maximize their shared entanglement.
Duality of entanglement sharability vs entanglement of assistance is evident in that
the upper bound for the former is the lower bound for the latter.

We use concurrence~\cite{Hil97,Woo98} for the entanglement measure amongst the
possible choices because of its simplicity and its appropriateness for  
distribution of entanglement~\cite{Gou04};  concurrence is generally used to study
MoE~\cite{Cof00} (although MoE
associated with von Neumann entropy is also studied~\cite{Koa04}).
When the N-partite state is pure and concurrence is the chosen entanglement
measure, this common bound is given by the linear entropy,
which is Dar\'{o}czy's $\beta$-entropy for $\beta=2$~\cite{Dar70}),
and arises often as a convenient form for, and estimator of, entropy~\cite{Ber03} .

In proving that entanglement of assistance is bounded below by the upper bound for
sharability of entanglement, we have established duality of sharability vs assistance of
entanglement. Furthermore, to prove this theorem, we introduce and prove 
bounds on the addition of a linear entropy version of quantum mutual information,
which yields new MoE inequalities despite linear entropy not being additive.

A consequence of our theorem is that we can bound total bipartite entanglement of
a pure~$N$-qubit state. Specifically our theorem leads to a useful corollary that proves 
that, for a four-qubit pure state, total entanglement for a bipartite cut with one qubit on one side and three qubits on the other always exceeds total entanglement for two qubits on each side of the cut, with a difference that is bounded by the minimum of the linear entropy for the individual qubits.
In the case of pure $N$-qubit states with~$N \geq 5$ qubits, we show that the total entanglement for two qubits on one side and the rest on the other side
can be bounded from above and below by functions corresponding to total
entanglement with one qubit on one side of the cut and the rest on the other side.

\section{Formulation}
\label{sec:formulation}

We are interested in an~$N$-qubit system whose state is described by the density operator
$\rho\in\mathcal{B}(\mathcal{H}_2^{\otimes N})$.
The joint density matrix for qubits~A and~B is given by
$\rho^\text{AB}=\text{Tr}_{\neq\text{AB}}\rho$, i.e.~the trace of the state over all qubits
except~A and~B, and the density matrix for qubit~A alone is
$\rho^\text{A}=\text{Tr}_{\neq\text{A}}\rho$.
We wish to discuss the shared information between the
pair of qubits~A and~B in terms of the
linear entropy (Dar\'{o}czy's $\beta=2$ entropy~\cite{Dar70}),
\begin{equation}
	S_\text{L}(\rho)\equiv 2(1-\text{Tr}\rho^2),
\end{equation}
for any density matrix~$\rho$ and therefore introduce the following form of
mutual information defined in terms of linear entropy.
\begin{definition}
	The linear mutual entropy between qubits~A and~B is
	\begin{equation}
		S_\text{L}(\text{A}:\text{B})\equiv S_\text{L}(\rho^\text{A})
			+S_\text{L}(\rho^\text{B})-S_\text{L}(\rho^\text{AB})\label{lme}
	\end{equation}
\end{definition}
This linear mutual entropy must be handled with care because, unlike von Neumann
entropy, linear entropy is not additive.
Hence `linear mutual entropy' is analogous
to mutual information but does not represent the standard meaning of mutual information.
For example, it does not vanish for product states.
However, as we shall see, linear mutual entropy is significant for bounding MoE.
One of our main results is tight lower and upper bounds on additivity of linear mutual entropy.

We will see throughout the paper that the linear mutual entropy is closely related to the 
concurrence~\cite{Woo98} and its dual, the concurrence of assistance~\cite{Lau01}. The concurrence
is a bipartite entanglement monotone quantifying the entanglement resources required to manufacture the
state~$\rho^\text{AB}$,
whereas the concurrence of assistance is a {\em tripartite} entanglement monotone~\cite{Gou05} 
quantifying the amount of bipartite entanglement that can be provided by the rest of the system
($\neq$AB) to establish maximum average concurrence between~A and~B. Both the concurrence and its 
dual are given in terms of the eigenvalues of the Hermitian operator,
\begin{equation}
	R^\text{AB}\equiv\sqrt{\sqrt{\rho^\text{AB}}\tilde{\rho}^\text{AB}\sqrt{\rho^\text{AB}}}\;,
\label{eq:R}
\end{equation}
where $\tilde{\rho}^\text{AB}$ is the spin-flipped density matrix,
\begin{equation}
	\tilde{\rho}^\text{AB}\equiv (\sigma_y\otimes\sigma_y)\rho^\text{AB*}(\sigma_y\otimes\sigma_y)\;,
\end{equation}
with $\rho^\text{AB*}$ the complex conjugate of $\rho^\text{AB}$ in the standard basis. 
Denoting by $\{\lambda^\text{AB}_1,\lambda^\text{AB}_2,\lambda^\text{AB}_3,\lambda^\text{AB}_4\}$
the four real nonnegative eigenvalues of $R^\text{AB}$ in decreasing order of magnitude, 
we have~\cite{Woo98,Lau01}
\begin{align}
& C^\text{AB} \equiv\text{min}\sum_ip_iC(|\psi_i\rangle^\text{AB}\langle\psi_i|)
= \text{max}\{0,\lambda^\text{AB}_1-\lambda^\text{AB}_2-\lambda^\text{AB}_3-\lambda^\text{AB}_4\}\label{eq:C}\\
& C_\text{a}^\text{AB}\equiv\text{max}\sum_ip_iC(|\psi_i\rangle^\text{AB}\langle\psi_i|)
=\text{Tr}[R^\text{AB}] = \sum_{i=1}^{4}\lambda_i^\text{AB}\;,\label{eq:CoA}
\end{align}
where the mimum and maximum in the definitions of $C^\text{AB}$ and $C_\text{a}^\text{AB}$
are taken over all decompositions of $\rho^\text{AB}=\sum_ip_i|\psi_i\rangle^\text{AB}\langle\psi_i|$.
For a pure state, $\rho^\text{AB}=(\rho^\text{AB})^2$, both the concurrence and its dual are given by
\begin{equation}
C^\text{AB}=C^\text{AB}_\text{a}=\sqrt{S_\text{L}(\rho^\text{AB})}\;.
\end{equation}

The contrast of the `max' in Eq.~(\ref{eq:CoA}) to the `min'
in Eq.~(\ref{eq:C}) underscores the duality between concurrence 
and concurrence of assistance. As we discuss now, this duality 
is manifested most elegantly in the MoE inequalities these measures 
satisfy.

First we discribe the MoE inequalities satisfied by the concurrence.
The tangle~\cite{Cof00} $\tau^\text{AB}\equiv (C^\text{AB})^2$ 
satisfies the original multi-qubit MoE inequality~\cite{Cof00,Osb05}
\begin{equation}
	\tau^{\text{A}\text{B}^{(1)}}+\tau^{\text{A}\text{B}^{(2)}}+\dots+\tau^{\text{A}\text{B}^{(N-1)}}
		\leq \tau^{(\text{A})(\text{B}^{(1)}\cdots\text{B}^{(N-1)})}
\label{eq:CKW}
\end{equation}
with~$N$ parties~A and~B$^{(1)},\ldots,$B$^{(N-1)}$,
and $\tau^{(A)(B\cdots B^{(N-1)})}$ represents the tangle between the bipartite split
of~A and~B$^{(1)}\cdots$B$^{(N-1)}$. Note that, although party~A appears to be privileged in these
analyses, the choice of~A from a network is arbitrary hence not really privileged:
any party could be chosen to be~A.
If the N-partite state is pure, 
then inequality~(\ref{eq:CKW}) can be simplified to
\begin{equation}
	\tau^{\text{A}\text{B}^{(1)}}+\tau^{\text{A}\text{B}^{(2)}}+\dots+\tau^{\text{A}\text{B}^{(N-1)}}
		\leq S_\text{L}(\rho^\text{A})\;.
\end{equation}

A dual version for monogamy of concurrence of assistance, given that the N-partite state is pure,
has been conjectured~\cite{Gou05}. We define tangle of assistance to be
\begin{equation}
	\tau_\text{a}^\text{AB}:=(C_\text{a}(\rho^\text{AB}))^2,
\end{equation}
and the conjecture follows~\cite{Gou05}.
\begin{conjecture}
Concurrence of assistance for a pure N-partite system satisfies the inequality
\begin{equation}
	S_\text{L}(\rho^\text{A})
		\leq \tau_\text{a}^{\text{A}\text{B}^{(1)}}
			+\tau_\text{a}^{\text{A}\text{B}^{(2)}}+\dots
			+\tau_\text{a}^{\text{A}\text{B}^{(N-1)}}.
\label{eq:conjecture}
\end{equation}
\end{conjecture}
Inequality~(\ref{eq:conjecture}) saturates for W~states~\cite{Gou05}. 
If the conjecture is true, it completes the monogamy inequalities for
pure N-partite states:
\begin{equation}
	\tau^{\text{A}\text{B}^{(1)}}+\tau^{\text{A}\text{B}^{(2)}}+\dots+\tau^{\text{A}\text{B}^{(N-1)}}
		\leq S_\text{L}(\rho^\text{A})\leq\tau_\text{a}^{\text{A}\text{B}^{(1)}}
			+\tau_\text{a}^{\text{A}\text{B}^{(2)}}+\dots
			+\tau_\text{a}^{\text{A}\text{B}^{(N-1)}}.
\end{equation}
We prove this conjecture here with the help of a powerful theorem in the next section
that provides tight lower and upper bounds on the sum of linear mutual entropies.

\section{Bounds for Additivity of Linear Mutual Entropy}

In this section we develop a theorem that provides lower and upper bounds on 
linear mutual entropy beginning with the following Lemma.
\begin{lemma}
The tangle of assistance provides an upper bound for the linear mutual entropy, namely
	\begin{equation}
		\tau_\text{a}^\text{AB}\geq \frac{1}{2}S_\text{L}(A:B).
	\end{equation}
\label{Lemma:ta}
\end{lemma}
\begin{proof}
This Lemma is easily proven by employing the result~\cite{Cof00}
\begin{equation}
\text{Tr}[(R^\text{AB})^2]
=1-\text{Tr}((\rho^\text{A})^2)-\text{Tr}((\rho^\text{B})^2)+\text{Tr}((\rho^\text{AB})^2)
=\frac{1}{2}S_\text{L}(A:B)\;,
\end{equation}
where the last equality follows from the definition of the linear mutual 
entropy~(\ref{lme}).
Thus, we have 
\begin{equation}
	\tau_\text{a}^\text{AB}
		= \left( \sum_{i-=1}^4 \lambda^\text{AB}_i \right)^2
		\ge \sum_{i=1}^4(\lambda^\text{AB}_i)^2
		= \text{Tr}[(R^\text{AB})^2]
		= \frac{1}{2}S_\text{L}(A:B)\;.
\label{mutual}
\end{equation}
\end{proof}
From the Lemma above it follows that
\begin{equation}
\frac{1}{2}\sum_{k=1}^{N-1}S_\text{L}(\text{A}:\text{B}^{(k)}) \leq  \sum_{k=1}^{N-1}
\tau_\text{a}\left(\rho^{\text{A}\text{B}^{(k)}}\right).
\end{equation}
Thus, in order to prove the conjecture given in Eq.~(\ref{eq:conjecture}), it is sufficient to prove that
\begin{equation}
	2S_\text{L}(\rho^\text{A})\leq\sum_{1=2}^{N-1} S_\text{L}(\text{A}:\text{B}^{(k)})\;.
\end{equation}
The following theorem establishes this lower bound, and therefore proves a stronger version
of the conjecture above, plus provides an upper bound. 
\begin{theorem}
	Let \(|\Psi\rangle\) be a pure N-qubit state. Then
\begin{enumerate}
\item[(a)] for~$N\geq 3$, $\;2S_L(\rho^\text{A})
	\leq\sum_{k=1}^{N-1}S_L(\text{A}:\text{B}^{(k)})\leq N S_L(\rho^\text{A})$, and
\item[(b)] for~$N=3,4$, $\sum_{k=1}^{N-1} S_L(\text{A}:\text{B}^{(k)})\leq (N-1) S_L(\rho^\text{A})$.
\end{enumerate}
\label{theorem:conj}
\end{theorem}
\begin{remark}
Due to strong subadditivity of the von Neumann entropy~\cite{Lie73},
part~(b) of Theorem~\ref{theorem:conj} would apply for all~$N\geq3$ if $S_\text{L}$ is
replaced by the von Neumann entropy. Whether
the LHS of part~(a) would be always satisfied if~$S_\text{L}$ is
replaced by the von Neumann entropy is left as an open problem.
\end{remark}

We divide the proof into two parts. In the first part we prove the left inequality
of~(a), from which the conjecture follows. Using similar techniques we also prove~(b). In the second
part we prove the right inequality of~(a) using a completely different approach. 

In the theorem above the~$N$-partite state is pure and the first qubit plays a special role,
although which qubit is decided to be first is discretionary and arbitrary. Therefore, as we will
see in the following, it would be helpful to present the theorem in a symmetric form, so that 
all qubits play the same role, which is used in
Lemmas~\ref{Lemma:D<}, \ref{Lemma:D>}, and~\ref{Lemma:D>>}.

As we henceforth assume that the N-partite state is pure, 
in order to present the theorem in a more symmetric form,
we write the pure state $\rho=|\Psi\rangle\langle|\Psi|$ in the following
Schmidt form:
\begin{equation}
	|\Psi\rangle=\sqrt{p_0}|0\rangle^\text{A}|\psi^{(0)}\rangle^\textbf{B}
		+\sqrt{p_1}|1\rangle^\text{A}|\psi^{(1)} \rangle^\textbf{B}
\label{schmidt}
\end{equation}
with $\{|0\rangle^\text{A},|1\rangle^\text{A}\}$ an orthonormal basis for the
first qubit, $|\psi^{(i)}\rangle^\textbf{B}$ two orthonormal states of the~$N-1$ other qubits,
and the superscript $^\textbf{B}$ referring to all~$N-1$ `B$^{(k)}$' qubits.
By denoting
\begin{equation}
	\sigma_k^{\ell\ell'}\equiv\text{Tr}_{\neq k}|\psi^{(\ell)}\rangle\langle\psi^{(\ell')}|, \; \ell,\ell'\in\{0,1\},
\label{eq:sigma_k}
\end{equation}
we obtain the single-qubit reduced states given by
\begin{equation}
	\rho^\text{A} 
		= \text{Tr}_{\neq\text{A}}|\Psi\rangle\langle\Psi|
		=p_0|0\rangle^\text{A}\langle 0|+p_1|1\rangle^\text{A}\langle 1|, \;
	\rho^{\text{B}^{(k)}}=\text{Tr}_{\neq\text{B}^{(k)}}|\Psi\rangle\langle\Psi|
		=p_0\sigma_k^{00}+p_1\sigma_k^{11},
\label{eq:singlequbitreduced}
\end{equation}
and the two-qubit reduced states are given by
\begin{equation}
\rho^{\text{A}\text{B}^{(k)}}=\text{Tr}_{\neq\text{A}\text{B}^{(k)}}|\Psi\rangle\langle\Psi|
	=\sum_{l,l'\in\{0,1\}}\sqrt{p_lp_{l'}}|l\rangle^\text{A} \langle l'|\otimes\sigma_k^{\ell\ell'}\;.
\label{eq:twoqubitreduced}
\end{equation}
By substituting these reduced densities matrices in the expressions for the linear entropy
and the linear mutual entropy that appear in Theorem~\ref{theorem:conj}, we obtain the following
Lemmas, which are equivalent to Theorem~\ref{theorem:conj}. 

Before stating the Lemmas, we defne the discriminant for the matrices~(\ref{eq:sigma_k}),
which simplifies the Lemmas.
\begin{definition}
	The discriminant of the~$k^\text{th}$ $\sigma$ matrix is
	\begin{equation}
		\mathfrak{D}_k^{(N)}\equiv\text{Tr}\left(\sigma_k^{00}\sigma_k^{11}
			-\sigma_k^{01}\sigma_k^{10}\right),
	\label{eq:Dk}
	\end{equation}
	and its sum is $\mathfrak{D}^{(N)}\equiv\sum_{k=2}^N\mathfrak{D}_k$.
\end{definition}
\begin{remark}
	The discriminant~$\mathfrak{D}^{(N)}$ is invariant under local unitary operations
	and can conveniently be expressed as a function of just two entanglement
	measures~\cite{Mey02,Wal04}. Furthermore, $\mathfrak{D}^{(N)}=0$ for GHZ-type 
	states and $\mathfrak{D}^{(N)}=N-2$ for $W$-type states.
\end{remark}

Using this notation, Theorem~\ref{theorem:conj} follows from the following three Lemmas.
\begin{lemma}
	$\mathfrak{D}^{(N)}\leq N-2$.
\label{Lemma:D<}
\end{lemma}
Lemma~\ref{Lemma:D<} is equivalent to the left-hand side of part~(a) of Theorem~\ref{theorem:conj}.
\begin{lemma}
	$\mathfrak{D}^{(N)}\geq 0,N=2,3$.
\label{Lemma:D>}
\end{lemma}
Lemma~\ref{Lemma:D>} is equivalent to part~(b) of Theorem~\ref{theorem:conj}.
\begin{lemma}
	$\mathfrak{D}^{(N)}\geq -1$.
\label{Lemma:D>>}
\end{lemma}
Lemma~\ref{Lemma:D>>} is equivalent to the right-hand side of part~(a) of Theorem~\ref{theorem:conj}.

Note that in the above Lemmas all qubits play the same role as we have traced
over the first qubit.

Now we proceed to prove Lemma~\ref{Lemma:D<}, which is a key result, and has
a rather long proof. To begin we introduce the following notations.
An~$N$-bit integer~$i\in\{0,1\}^N$ can be expressed as
$i=\sum_{k=0}^{N-1}2^{k}i_k$ and represented by the bit sequence
$\vecb{i}\equiv (i_0,i_1,\ldots,i_{N-1})$. In dealing with the~$k^\text{th}$ qubit,
we need to be able to selectively modify the~$k^\text{th}$ bit in the sequence
and therefore introduce the notations
\begin{equation}
	\bar{\vecb{i}}_k\equiv(i_0,i_1,\ldots ,i_{k-1},\bar{i}_k,i_{k+1},\ldots ,i_{N-1}),\;\text{and}\;
	\vecb{i}_k^\ell\equiv(i_0,i_1,\ldots ,i_{k-1},\ell,i_{k+1},\ldots ,i_{N-1}),\; \ell\in\{0,1\},
\end{equation}
with $\bar{i}_k\equiv i_k+1\bmod 2$.

Two integers $i,j\in\{0,1\}^N$, are separated by the Hamming distance
$\Delta_{ij}=\sum_{k=0}^{N-1}|i_k-j_k|$. Sometimes we sum over all integer
pairs $(i,j)$ such that their Hamming distance is fixed to~$\Delta$. Such a sum
will be expressed as
\begin{equation}
	\sum_{\Delta_{ij}=\Delta}\equiv\sum_{i,j\in\{0,1\}^N}\delta_{\Delta,\Delta_{ij}},
\label{eq:Delta_ij}
\end{equation}
and the set of indices for which the two integers~$i$ and~$j$ differ is
\begin{equation}
	S_{ij}\equiv\{k;i_k=\bar{j}_k\}
\label{eq:S_ij}
\end{equation}
with cardinality~$\Delta_{ij}$.
For given~$i,j$, we define a set of pairs of bit strings
\begin{equation}
	G_{ij}=\left\{(i',j');S_{ij}=S_{i'j'}, i_k=i'_k \text{ and }j_k=j'_k
		\forall\, k\notin S_{ij}\right\}.
\end{equation}
with cardinality~$2^{\Delta_{ij}}$.
As $i_k=j_k$ for $k\notin S_{ij}$, $(i',j')\in G_{ij}\iff (j',i')\in G_{ij}$.

\begin{proof} (of Lemma~\ref{Lemma:D<})
The two orthonormal states for the \textbf{B} parties~(\ref{schmidt}) can be expressed as
$|\psi^{(\ell)}\rangle=\sum_{i\in\{0,1\}^N}a_i^{(\ell)}|i\rangle$, $\ell\in\{0,1\}$,
with
\begin{equation}
	\sum_i a_i^{(\ell)*}a_i^{(\ell')}=\delta_{\ell\ell'}.
\label{eq:orthonormality}
\end{equation}
Then we obtain
\begin{equation}
	\sigma_k^{\ell\ell'}=\sum_{\vecb{i}}
		\begin{pmatrix}
			a_{\vecb{i}_k^0}^{(\ell)} a_{\vecb{i}_k^0}^{(\ell')*}& 
			a_{\vecb{i}_k^0}^{(\ell)} a_{\vecb{i}_k^1 }^{(\ell')*}	\\
			a_{\vecb{i}_k^1}^{(\ell)}a_{\vecb{i}_k^0}^{(\ell')*}& 
			a_{\vecb{i}_k^1}^{(\ell)}b_{\vecb{i}_k^1 }^{(\ell')*}
		\end{pmatrix},
\label{eq:sigmamatrix}
\end{equation}
which is substituted into the expression for the discriminant, yielding
\begin{align}
	\mathfrak{D}^{(N)}= & \sum_{k=0}^{N-1}\sum_i\sum_j
	\Big(|a_{\vecb{i}_k^0}^{(0)}|^2|a_{\vecb{j}_k^0}^{(1)}|^2
		+|a_{\vecb{i}_k^1 }^{(0)}|^2|a_{\vecb{j}_k^1}^{(1)}|^2
		-a_{\vecb{i}_k^0}^{(0)} a_{\vecb{i}_k^0}^{(1)*}a_{\vecb{j}_k^0}^{(0)*}a_{\vecb{j}_k^0}^{(1)}
		-a_{\vecb{i}_k^1}^{(0)} a_{\vecb{i}_k^1}^{(1)*}a_{\vecb{j}_k^1}^{(0)*}a_{\vecb{j}_k^1}^{(1)}
			\nonumber \\
		&+a_{\vecb{i}_k^0}^{(0)} a_{\vecb{i}_k^1 }^{0*}a_{\vecb{j}_k^1}^{(1)}a_{\vecb{j}_k^0}^{(1)*}
		+a_{\vecb{i}_k^1}^{(0)}a_{\vecb{i}_k^0}^{(0)*}a_{\vecb{j}_k^0}^{(1)}a_{\vecb{j}_k^1}^{(1)*}
		-a_{\vecb{i}_k^0}^{(0)}a_{\vecb{i}_k^1}^{(1)*}a_{\vecb{j}_k^0}^{(0)*}a_{\vecb{j}_k^1}^{(1)}
		-a_{\vecb{i}_k^1}^{(0)}a_{\vecb{i}_k^0}^{(1)*}a_{\vecb{j}_k^1}^{(0)*}a_{\vecb{j}_k^0}^{(1)}\Big)
			\nonumber \\
		=&\sum_{\Delta=0}^N(N-\Delta)\sum_{\Delta_{ij}=\Delta}
			\left(|a_{\vecb{i}}^{(0)}|^2|a_{\vecb{j}}^{(1)}|^2
			-a_{\vecb{i}}^{(0)}a_{\vecb{i}}^{(1)*}a_{\vecb{j}}^{(0)*}a_{\vecb{j}}^{(1)}\right)\nonumber \\
			\nonumber \\
		&+\sum_{k=0}^{n-1}\sum_i\sum_j \left(
	a_{\vecb{i}_k^0}^{(0)}a_{\vecb{j}_k^1}^{(1)}\left[a_{\vecb{i}_k^1}^{(0)*}a_{\vecb{j}_k^0}^{(1)*}
	-a_{\vecb{i}_k^1 }^{(1)*}a_{\vecb{j}_k^0}^{(0)*}\right]
	+a_{\vecb{i}_k^1}^{(0)}a_{\vecb{j}_k^0}^{(1)}\left[a_{\vecb{i}_k^0}^{(0)*}a_{\vecb{j}_k^1 }^{(1)*}
	-a_{\vecb{i}_k^0}^{(1)*}a_{\vecb{j}_k^1}^{(0)*}\right]\right).
\label{eq:Dsum}
 \end{align}
Introducing
 \begin{equation}
 	\alpha_{\vecb{i}\vecb{j}}
		\equiv a_{\vecb{i}}^{(0)}a_{\vecb{j}}^{(1)}-a_{\vecb{j}}^{(0)}a_{\vecb{i}}^{(1)}
		= -\alpha_{\vecb{j}\vecb{i}},
\end{equation}
simplifies Eq.~(\ref{eq:Dsum}) to
\begin{equation}
	\mathfrak{D}^{(N)}
 		=\frac{1}{2}\sum_{\Delta=0}^N(N-\Delta)\sum_{\Delta_{ij}=\Delta}|\alpha_{\vecb{i}\vecb{j}}|^2
		 	+\sum_{k=0}^{N-1}\sum_i\sum_j\alpha_{\vecb{i}_k^0\vecb{j}_k^1 }
	\alpha_{\vecb{i}_k^1 \vecb{j}_k^0}^*.
\label{eq:DNalpha}
\end{equation}
The second term on the right-hand side of Eq.~(\ref{eq:DNalpha}) can be expressed as
\begin{equation}
	\sum_{\Delta=0}^N\Lambda_\Delta
\end{equation}
with
\begin{equation}
	\Lambda_\Delta\equiv\frac{1}{2}\sum_{\Delta_{ij}=\Delta}\alpha_{\vecb{i}\vecb{j}}
		\sum_{k\in S_{ij}}\alpha_{\vecb{i}_k\vecb{j}_k}^*.
\label{eq:2ndterm}
\end{equation}
In this equation, the term corresponding to $\Delta=0$ is zero because
$\vecb{i}=\vecb{j}\implies\alpha_{\vecb{i}\vecb{j}}=0$, and
the term corresponding to $\Delta=1$ equals
$-\tfrac{1}{2}\sum_{\Delta_{ij}=1}|\alpha_{\vecb{i}\vecb{j}}|^2$
because, in this case, the set $S_{ij}$ contains only one index, 
say~$k$, and, therefore, $\vecb{i}_k=\vecb{j}$ and $\vecb{j}_k=\vecb{i}$.
The term corresponding to $\Delta=2$ in Eq.~(\ref{eq:2ndterm}) is zero
because the set $S_{ij}$ contains exactly
two indices, $k_1$ and $k_2$, and, from the definition of $S_{ij}$,
$\vecb{i}_{k_1}=\vecb{j}_{k_2}$ and $\vecb{j}_{k_1}=\vecb{i}_{k_2}$.
Thus, in this case $\sum_{k\in S_{ij}}\alpha_{\vecb{i}_k\vecb{j}_k}^*=0$.
Collecting all these results yields
\begin{equation}
	\mathfrak{D}^{(N)} =\frac{1}{2}(N-2)\sum_{\Delta_{ij}
		=1}|\alpha_{\vecb{i}\vecb{j}}|^2+\frac{1}{2}
		\sum_{\Delta=2}^N(N-\Delta)\sum_{\Delta_{ij}
		=\Delta}|\alpha_{\vecb{i}\vecb{j}}|^2\nonumber \\
 		+\sum_{\Delta=3}^N\Lambda_{\Delta}.
\label{aaa}
\end{equation}
Orthonormality~(\ref{eq:orthonormality}) implies that $\frac{1}{2}\sum_{i,j}|\alpha_{ij}|^2=1$; hence
Eq.~(\ref{aaa}) can be rewritten as
\begin{equation}
	\mathfrak{D}^{(N)}=N-2+\sum_{\Delta
	=3}^N\left[\Lambda_{\Delta}+\frac{2-\Delta}{2}\sum_{\Delta_{ij}
	=\Delta}|\alpha_{\vecb{i}\vecb{j}}|^2\right].
\label{exp}
\end{equation}
Now, to prove the Lemma, we just need to show that
\begin{equation}
	\Lambda_{\Delta}\leq\frac{1}{2}(\Delta-2)\sum_{\Delta_{ij}
	=\Delta}|\alpha_{\vecb{i}\vecb{j}}|^2.
\label{eq:LambdaDelta}
\end{equation}
The sum in the expression for~$\Lambda_\Delta$~(\ref{eq:2ndterm}) can be written as
\begin{equation}
	\Lambda_{\Delta}\equiv\frac{1}{2}\sum_{\{ G_{ij}\}}\sum_{(\vecb{i}',\vecb{j}')\in G_{ij}^	{\Delta}}\sum_{k\in S_{ij}}\alpha_{\vecb{i}'\vecb{j}'}\alpha_{\vecb{i}_k'\vecb{j}_k'}^*
\end{equation}
with the first sum taken over all distinct sets $G_{ij}$ with identical~$\Delta$.

Given a set $G_{ij}$, if $(i',j')\in G_{ij}$
then $\vecb{j}'$ is determined uniquely by $\vecb{i}'$. Moreover,
by definition, since $(\vecb{i}',\vecb{j}')\in G_{ij}$ we
have $i_k=i'_k$ for $k\notin S_{ij}$. That is, the pair $(i',j')$
is determined uniquely by the string of $\Delta$ bits $\{ i_k'\}$
with $k\in S_{ij}$. We denote this string of $\Delta$ bits by $\vecb{x}=(x_0,x_1,\ldots ,x_{\Delta-1})$.
Note that there is a one-to-one correspondence between the $2^{\Delta}$
pairs $(i',j')\in G_{ij}$ and the $2^{\Delta}$
strings $\{\vecb{x}\}$. Using this correspondence, we denote
\begin{equation}
	\alpha_{\vecb{x}}\equiv\alpha_{\vecb{i}'\vecb{j}'}.
\end{equation}
Now, since $\alpha_{\vecb{i}'\vecb{j}'}=-\alpha_{\vecb{j}'\vecb{i}'}$, we have 
\begin{equation}
	\alpha_{\vecb{x}}=-\alpha_{\vecb{y}}\label{ij}
\end{equation}
for 
\begin{equation}
	\vecb{y}=(\bar{x}_0,\bar{x}_1,\ldots ,\bar{x}_{\Delta-1}).
\end{equation}

Thus, with this notation we can write
\begin{equation}
	\frac{1}{2}\sum_{(\vecb{i}',\vecb{j}')\in G_{ij}}
	\sum_{k\in S_{ij}}\alpha_{\vecb{i}'\vecb{j}'}\alpha_{\vecb{i}_k'\vecb{j}_k'}^*
	=\frac{1}{2}\sum_{x_0=0}^1 \cdots\sum_{x_{\Delta-1}=0}^1 
	\alpha_{\vecb{x}}\sum_{k=0}^{\Delta-1}\alpha_{\vecb{x}_k}^*,
\end{equation}
where $\vecb{x}_k=(x_0,x_1,\ldots ,x_{k-1},\bar{x}_k,x_{k+1},\ldots ,x_{\Delta-1})$.
Therefore, in order to prove Eq.~(\ref{eq:LambdaDelta}), it is sufficient to show that
\begin{equation}
	\frac{1}{2}\sum_{\vecb{x}} \alpha_{\vecb{x}}
	\sum_{k=0}^{\Delta-1}\alpha_{\vecb{x}_k}^*
	\leq\left(\frac{\Delta-2}{2}\right)\sum_{\vecb{x}}|\alpha_{\vecb{x}}|^2.
\label{eq:alphax}
\end{equation}
Let us denote by $\vecb{\alpha}$ a vector
with the $2^{\Delta-1}$ components $\{\alpha_{\vecb{x}}^*\}_{\vecb{x}=(000\ldots 0)}^{(011\ldots 1)}$
(i.e. $x_0$ is kept zero). Note that due to Eq.~(\ref{ij}), the
RHS of Eq.~(\ref{eq:alphax}) is equal to $(\Delta-2)||\vecb{\alpha}||^2$.
Moreover, also the LHS can be written in a compact form and from Eq.~(\ref{ij}),
it follows that the inequality given in Eq.~(\ref{eq:alphax}) is equivalent
to \begin{equation}
\vecb{\alpha}^{\dag}V\vecb{\alpha}\leq(\Delta-2)||\vecb{\alpha}||^2\end{equation}
 where the $2^{\Delta-1}\times2^{\Delta-1}$ matrix $V$ is given
by
\begin{equation}
	V_{xy}=\left\{ 
		\begin{array}{rll}
			-1&\text{if }x_k=\bar{y}_k&\text{for all }k=1,2,\ldots,\Delta-1\\
			1&\text{if }x_k=y_k&\text{for exactly }\Delta-2\;\; k{\text{'s}}\\
			0&&\text{otherwise}
		\end{array}
	\right.
\end{equation}
 where $x,y\in\{0,1,2,\ldots ,2^{\Delta-1}-1\}$ corresponds to
 \begin{equation}
 	(x_1,x_2,\ldots ,x_{\Delta-1}),\;(y_1,y_2,\ldots ,y_{\Delta-1}),
\end{equation}
respectively. 
Now, in order to prove Eq.~(\ref{eq:alphax}), it is left to show that the largest eigenvalue
of the matrix $V$ is $\Delta-2$. In order to show that we define
the $2^m\times2^m$ matrices $P_m$. The definition is by induction:
\begin{equation}
P_1=\begin{pmatrix}1 & 1\cr1 & -1\cr\end{pmatrix}\;\;{\text{and}}\;\; 
P_m=\begin{pmatrix}P_m & P_m\cr P_m & -P_m\cr\end{pmatrix}.
\end{equation}
 It is easy to see from the definition that $P_m^2=2^mI$, where
$I$ is the $2^m\times2^m$ identity matrix. This implies that
the columns of $P_m$ are linearly independent. The components of
the $2^m\times2^m$ matrix $P_m$ are given by
\begin{equation}
	\left(P_m\right)_{x,y}=(-1)^{\vecb{x}\cdot\vecb{y}},\;\;\; x,y\in\{0,1,\ldots ,2^m-1\}
\end{equation}
where $\vecb{x}$ and~$\vecb{y}$ are the strings of bits corresponding
to~$x$ and~$y$, respectively.

With this explicit expression for $P_m$, it is a simple exercise
to check that the the columns of $P_{\Delta-1}$ form a basis of eigenvectors
of $V$. Using these $2^{\Delta-1}$ eigenvectors, we find that the
eigenvalues of $V$ are given by: \begin{equation}
\lambda_{y}=\sum_{k=1}^{\Delta-1}(-1)^{y_k}-(-1)^{\sum_{k=1}^{\Delta-1}y_k}\;.\end{equation}
 As all $y_k\in\{0,1\}$, it follows that $\lambda_{y}\leq\Delta-2$, and this completes 
 the proof of Lemma~\ref{Lemma:D<}.
 \end{proof}

We now prove Lemma~\ref{Lemma:D>}. 
\begin{proof}
	(of Lemma~\ref{Lemma:D>}).

For the case of~$N=3$ qubits, it is easy to verify that
$\mathfrak{D}^{(N)}=0$. We now show that, for~$N=4$ qubits,
we have $\mathfrak{D}^{(4)}\geq 0$. In fact, we find an explicit formula
for $\mathfrak{D}^{(4)}$. We start by writing Eq.~(\ref{exp}) for four qubits
($N=3$) as
\begin{equation}
	\mathfrak{D}^{(4)}=1+\frac{1}{2}\sum_{\Delta_{ij}=3}\left(\alpha_{\vecb{i}\vecb{j}}\sum_{k\in 
	S_{ij}}\alpha_{\vecb{i}_k\vecb{j}_k}^*-|\alpha_{\vecb{i}\vecb{j}}|^2\right)
\end{equation}
where we have used the definition of $\Lambda_\Delta$ in Eq.~(\ref{eq:2ndterm}).
This expression can be written explicitly in the following form: 
\begin{equation}
	\mathfrak{D}^{(4)}=1-|\alpha_{000,111}-\alpha_{001,110}-
	\alpha_{010,101}+\alpha_{011,100}|^2.
\end{equation}
In terms of $a_{\vecb{i}}$ and $b_{\vecb{i}}$ it is given by the following expression:
\begin{equation}
%\begin{align}
	\mathfrak{D}^{(4)}  =1-\Big|a_{000}^{(0)}a_{111}^{(1)}+a_{110}^{(0)}a_{001}^{(1)}
	+a_{101}^{(0)}a_{010}^{(1)}+a_{011}^{(0)}a_{100}^{(1)}%\nonumber \\
	 -a_{111}^{(0)}a_{000}^{(1)}-a_{001}^{(0)}a_{110}^{(1)}
	-a_{010}^{(0)}a_{101}^{(1)}-a_{100}^{(0)}a_{011}^{(1)}\Big|^2.
%\end{align}
\end{equation}
Thus, from the above expression and Eq.~(\ref{eq:orthonormality}),
it follows that $\mathfrak{D}^{(4)}\geq0$.
\end{proof}

Before we proceed to the proof of Lemma \ref{Lemma:D>>} which rather uses a 
different method, we first prove an important monogamy inequality involving 
tangle and tangle of assistance. We begin with some useful expressions.
From Eq.~(\ref{mutual}), the tangle of assistance is
\begin{equation}
\tau_\text{a}^\text{AB} = \left(\sum\lambda^\text{AB}_i\right)^2 
	= \frac{1}{2}S_\text{L}(\text{A}:\text{B})+X^\text{AB}+Y^\text{AB}\;,
\label{calT}
\end{equation}
where
\begin{equation}
	X^\text{AB} = 2\sum_{k=2,3,4}\lambda_1^\text{AB}\lambda_k^\text{AB}, \;\text{and}\;\;
	Y^\text{AB} = 2\sum_{k<l=2,3,4}\lambda_k^\text{AB}\lambda_l^\text{AB}\;.
\end{equation}
For clarity, henceforth we omit the superscript \text{AB} for $\lambda$. 

The tangle can also be expressed in a similar way. If $\lambda_1 \leq \lambda_2+\lambda_3+\lambda_4$ then
$\tau^\text{AB}=0$; otherwise
\begin{equation}
\tau^\text{AB}  = (\lambda_1-\lambda_2
	-\lambda_3-\lambda_4)^2
 = \frac{1}{2}S_\text{L}(\text{A}:\text{B})-X^\text{AB}+Y^\text{AB}
\end{equation}

Note that if $\tau^\text{AB}>0$ (i.e. $\lambda_1 > \lambda_2+\lambda_3+\lambda_4$) then
$\tau_\text{a}^\text{AB}+\tau^\text{AB}=
S_\text{L}(\text{A}:\text{B})+2Y^\text{AB}\geq S_\text{L}(\text{A}:\text{B})$. On the otherhand,
if $\lambda_1 \leq \lambda_2+\lambda_3+\lambda_4$ (i.e. $\tau^\text{AB}=0$) then
it is easy to check that $\lambda_1^2+\lambda_2^2+\lambda_3^2+\lambda_4^2\leq X^\text{AB}+Y^\text{AB}$;
that is, $\tau_\text{a}^\text{AB}\geq S_\text{L}(\text{A}:\text{B})$. We summerize this observation
in the following claim.
\begin{claim} 
\begin{equation}
	\tau_\text{a}^\text{AB}+\tau^\text{AB}\geq S_\text{L}(\text{A}:\text{B})\;.
\end{equation}
\label{claim:D>>>}
\end{claim}
We are now ready for the proof of of Lemma~\ref{Lemma:D>>}.

\begin{proof}
	(of Lemma~\ref{Lemma:D>>}).

In~\cite{Gou05} it has been shown that for any choice of entanglement monotone $E$,
the entanglement of assistance, $E_\text{a}(\rho^\text{AB})$, is bounded from above by 
${\rm min}\{E^{\text{A}({\rm rest})},E^{\text{B}({\rm rest})}\}$, where $E^{i({\rm rest})},i=\text{A,B}$ is the bipartite entanglement shared between qubit $i$ and the rest of the system. 
Taking $E$ to be the concurrence gives
\begin{equation}
	C_\text{a}^\text{AB}\leq{\rm min}\left\{\sqrt{S_\text{L}(\rho^\text{A})},\sqrt{S_\text{L}(\rho^\text{B})}\right\},
\end{equation}
or, equivalently,
\begin{equation}
	\tau_\text{a}^\text{AB}\leq\min\{ S_\text{L}(\rho^\text{A}),S_\text{L}(\rho^\text{B})\}.
\end{equation}
Thus we have
\begin{equation}
	\sum_{i=1}^{N-1}\tau_\text{a}^{\text{A}\text{B}^{(i)}}\leq(N-1)S_\text{L}(\rho^\text{A}).
\label{equation:sumass}
\end{equation}
From the CKW inequality,
\begin{equation}
	\sum_{i=1}^{N-1}\tau^{\text{A}\text{B}^{(i)}}\leq S_\text{L}(\rho^\text{A}).
\label{equation:sumtan}
\end{equation}
Combining Eqs.~(\ref{equation:sumass}) and~(\ref{equation:sumtan}) yields 
\begin{equation}
	\sum_{i=1}^{N-1}\left(\tau_\text{a}^{\text{A}\text{B}^{(i)}}+\tau^{\text{A}\text{B}^{(i)}}\right)\leq N S_\text{L}(\rho^\text{A}).
\end{equation}
Finally, using Claim~\ref{claim:D>>>} we find
\begin{equation}
	\sum_{i=1}^{N-1} S_\text{L}(\text{A}:\text{B}^{(i)})\leq\sum_{i=1}^{N-1}\left(\tau_\text{a}^{\text{A}\text{B}^{(i)}}+\tau^{\text{A}\text{B}^{(i)}}\right)\leq NS_\text{L}(\rho^\text{A}),
\end{equation}
thereby proving Lemma~\ref{Lemma:D>>} which is equivalent to the right hand side of part (a) of Theorem~\ref{theorem:conj},
\end{proof}

From Claim~\ref{claim:D>>>} it follows that the tangles (and their dual)
of three parties satisfy 
\begin{equation}
	\tau^\text{AB}+\tau^\text{AC}+\tau_\text{a}^\text{AB}+\tau_\text{a}^\text{AC}
	\geq S_\text{L}(\text{A}:\text{B})+S_\text{L}(\text{A}:\text{C}).
\end{equation}
If the parties sharing a pure 3-qubit state then it can be easily shown that this inequality
becomes the following equality:
\begin{equation}
	\tau^\text{AB}+\tau^\text{AC}+\tau_\text{a}^\text{AB}+\tau_\text{a}^\text{AC}
	=2S_\text{L}(\rho^\text{A}).
\end{equation}
The above equality leads
to an interesting consequence.
For any three party pure state~\cite{Koa04}
\begin{equation}
	E_\text{f}(\rho^\text{AB})+I^{\leftarrow}(\rho^\text{AC})=S(\rho^\text{A})\;,
\end{equation}
where $E_\text{f}(\cdot)$ is the entanglement of formation, $I^{\leftarrow}(\cdot)$
is the one way classical correlation measure first introduced by Henderson
and Vedral~\cite{Hen01} (see also Devetak and Winter~\cite{Dev03} for the operational interpretation
of this measure) and 
$S(\rho^\text{A})$ is the von Neumann entropy of $\rho^\text{A}$.
Osborne and Verstrate~\cite{Osb05} further showed that for three qubits, the above
equality is valid if we replace the von Neumann entropy that appears
in the definitions of $E_\text{f}$ and $I^{\leftarrow}$ with the linear entropy
and also in the right hand side. In particular,
\begin{equation}
\tau(\rho^\text{AB})+I^{\leftarrow}{}_l (\rho^\text{AC})=S_\text{L}(\rho^\text{A})\;.
\end{equation}
Evidently, for three qubit pure states,
\begin{equation}
\tau(\rho^\text{AB})+\tau(\rho^\text{AC})+I^{\leftarrow}{}_l (\rho^\text{AB})+I^{\leftarrow}{}_l 
(\rho^\text{AC})=2S_\text{L}(\rho^\text{A}).
\end{equation}
 Thus we have established the following equality for three qubit pure states
\begin{equation}
I^{\leftarrow}{}_l (\rho^\text{AB})+I^{\leftarrow}{}_l (\rho^\text{AC})=\tau_\text{a}(\rho^\text{AB})
+\tau_\text{a}(\rho^\text{AC}).
\end{equation}
 The above equality is particulary interesting as the left hand side represents total one way classical 
correlation (von Neumann entropy replaced by linear entropy) between the pairs of
qubits AB and AC whereas the
right hand side is purely quantum; it implies a possible connection between distilable common 
randomness~\cite{Dev03} and entanglement of assistance. 

\section{Implications}
Theorem~\ref{theorem:conj} proves the conjecture and also provides new bounds on total entanglement (measured by the
tangle) across various bipartitions and entanglement relationships among the different possible bipartite cuts of 
multipartite systems. 
Now suppose we have a N-partite pure state and we wish to know the ordering of total bipartite entanglement across the 
bipartitions. 

For notational convenience in this Section, we introduce $\text{C}^{(1)}\equiv\text{A}$ and
$\text{C}^{(k)}\equiv\text{B}^{(k-1)}$ for 
$\text{k}=2,\ldots,N+1$.  For any multipartite~$N$-qubit pure state, $\tau^{k:{\text{rest}}}$ is the 
tangle between qubit $C^{(k)}$ and the rest and $\tau^{kj:{\text{rest}}}$ is the 
tangle between the pair of qubits $C^{(k)}$ and~$C^{(j)}$ and the rest.  (Note that 
expressions for these tangles are equivalent under an interchange of~$k$ and~$j$.)

One interesting question is whether total entanglement  across some 
bipartition dividing one qubit from the rest, i.e.~a `$k:\text{rest}$' bipartition, which puts qubit~$k$ on
one side and the other~$N-1$ qubits on the other side,
is greater than or less than the total entanglement across some $kk':\text{rest}$ bipartition,
which puts both qubits~$k$ and~$k'$ on one side and the rest of the qubits on the other side.
In other words one would like to know quantitative relationships between $k:\text{rest}$ and 
$kk':\text{rest}$ bipartitions.
Fortunately Theorem~\ref{theorem:conj} provides answers to some of these questions. 

For~$N=3$, a $k:\text{rest}$ bipartition is equivalent to a $kk':\text{rest}$ bipartition 
so the comparison between the two is trivial. Therefore, we restrict to~$N\geq 4$ for non-trivial cases.

In order to relate entanglement between bipartitions of type $k:\text{rest}$ vs $kk':\text{rest}$,
we compute and compare a weighted sum of tangles across \emph{all} bipartitions that are of the form
$k:\text{rest}$ vs the sum of the tangles across \emph{all} bipartitions that are of the form
$kk':\text{rest}$. These two raw sums are unequally weighted;
that is, for large~$N$, there are many more $kk':\text{rest}$ bipartitions, which scales as~$N^2$,
than there are~$k:\text{rest}$ bipartitions (scales as~$N$).
Thus, we would like first to compare between quantities with 
the same weight and therefore start with the following definitions. 

\begin{definition}
	Total entanglement across all bipartitions of the form $k':\text{rest}$ with the~$k^{\text{th}}$ qubit amidst the `rest'
	(i.e.\ does not appear alone) is denoted by 
	\begin{equation}
		\tau_1^k \equiv \sum_{\ell\neq k}\tau^{\ell:\text{rest}}\;.
	\end{equation}
\end{definition}
\begin{definition}
	Total entanglement across all bipartitions of the form $k'k'':\text{rest}$ with the~$k^{\text{th}}$ qubit appears on the side with exactly one more qubit is denoted by 
	\begin{equation}
		\tau_2^k \equiv \sum_{\ell\neq k}\tau^{k\ell:\text{rest}}.
	\end{equation}
\end{definition}
Note that the sums in the definitions of $\tau_1^k$ and $\tau_2^k$ each comprise~$N-1$ terms. 

Our next result  provides both lower and upper bounds for $\tau_2^k$ in terms
of $\tau_1^k$ and $\tau^{k:\text{rest}}$.
\begin{corollary}
For any multipartite~$N$-qubit pure state, 
\begin{equation}
	(\delta_{N,4}-1)\tau^{k:\text{rest}}\leq\tau_2^k-\tau_1^k\leq (N-3) \tau^{k:\text{rest}}.
\label{bounds}
\end{equation}
\label{corollary:tau-k:rest}
\end{corollary}
Note that for~$N=4$ the lower bound for $\tau_2^k-\tau_1^k$ is 
zero~\footnote{For the case of four qubits we were able to prove a slightly tighter bound 
-see Theorem~\ref{theorem:conj}.}. Due to the strong subadditivity of the von Neumann 
entropy, this would be the lower bound for all~$N\geq 4$ if the entanglement is measured by 
the \emph{entropy of entanglement} instead of the tangle (the question regarding the upper bound,
however, is left open).
\begin{proof}
(of Corollary~\ref{corollary:tau-k:rest}).
We begin with the the result of Theorem~\ref{theorem:conj}, namely
\begin{equation}
2S_L(\rho^\text{A})\leq\sum_{j=1}^{N-1}S_L(\text{A}:\text{B}^{(j)})\leq (N-\delta_{N,4})S_L(\rho^\text{A})\;.
\label{start}
\end{equation}
Thus Eq.~(\ref{start}) in the new notation is given by 
\begin{equation}
	2S_L(\rho^{\text{C}^{(1)}})\leq\sum_{j=2}^NS_L(\text{C}^{(1)}:\text{C}^{(j)})
		\leq (N-\delta_{N,4})S_L(\rho^{\text{C}^{(1)}})\;.
\label{startnew}
\end{equation}
Expanding~(\ref{startnew}) yields
\begin{equation}
	2S_L(\rho^{\text{C}^{(1)}})\leq(N-1)S_L(\rho^{\text{C}^{(1)}})
		+\sum_{j=2}^NS_L(\rho^{\text{C}^{(j)}})
		-\sum_{j=2}^NS_L(\rho^{\text{C}^{(1)}\text{C}^{(j)}})\leq NS_L(\rho^{\text{C}^{(1)}}).
\end{equation}
From the left-hand inequality,
\begin{equation}
	\sum_{j=2}^NS_L(\rho^{\text{C}^{(1)}\text{C}^{(j)}}) \leq  (N-3)S_L(\rho^{\text{C}^{(1)}})
		+ \sum_{j=2}^NS_L(\rho^{\text{C}^{(j)}})\;,
\end{equation}
and from the right-hand inequality,
\begin{equation}
	\sum_{j=2}^NS_L(\rho^{\text{C}^{(j)}})-(1-\delta_{N,4})S_L(\rho^{\text{C}^{(1)}})
		\leq\sum_{j=2}^NS_L(\rho^{\text{C}^{(1)}\text{C}^{(j)}})\;.
\end{equation}
Thus, in terms of the tangle $\tau^{k:\text{rest}}\equiv S_L(\rho^{\text{C}^{(k)}})$ and
$\tau^{(1k:\text{rest})}\equiv S_L(\rho^{\text{C}^{(1)}\text{C}^{(k)}})$, the inequalities above can be 
written as 
\begin{equation}
	(\delta_{N,4}-1)\tau^{1:\text{rest}}\leq\sum_{k=2}^N\tau^{1k:\text{rest}}
		-\sum_{k=2}^N \tau^{k:\text{rest}}\leq(N-3)\tau^{1:\text{rest}}.
\label{ent5}
\end{equation}
In Eq.~(\ref{ent5}), the first qubit plays a special role, although which qubit is decided to
be first is discretionary and arbitrary. Hence Eq.~(\ref{ent5}) is equivalent to Eq.~(\ref{bounds}).
\end{proof}

Note that if we sum over all $k$ in Eq.~(\ref{bounds}), we get
\begin{equation}
\frac{N-2+\delta_{N,4}}{2}\tau_1\leq\tau_2\leq (N-2) \tau_1
\end{equation}
with 
\begin{equation}
	\tau_1\equiv\sum_{k=1}^N\tau^{k:\text{rest}},\;
	\tau_2\equiv\tfrac{1}{2}\sum_{k=1}^N\tau^{k}_2
		=\sum_{\stackrel{k,k'=1}{j<k}}^N\tau^{(kk':\text{rest})}\;.
\end{equation}

That is, we have found both lower and upper bounds for total entanglement across all 
bipartite cuts of the type $kk':\text{rest}$ in terms of total entanglement across all bipartitions of
the type $k:\text{rest}$.

\section{Conclusions}

Entanglement is a key resource for quantum information processing and is a quintessential feature of quantum physics, yet relationships for multipartite entanglement are still not well understood. One of the most important multipartite entanglement relations is monogamy, which bounds the amount of entanglement that can be shared in a multipartite system. The Coffman-Kundu-Wootters conjecture~\cite{Cof00}, only recently proven by Osborne and Verstraete~\cite{Osb05}, captures the essence of monogamy of entanglement in terms of concurrence, and tangle provides an upper bound to the sharing of entanglement in a multi-partite system.

Our interest is in concurrence of assistance, which expresses how much concurrence can be shared by two parties in the network through the assistance of another party or the rest. In contrast to monogamy of entanglement, which corresponds to an upper bound on the sharability of entanglement, we have shown that same bound is a \emph{lower bound} on how much entanglement can be created by assistance, as measured by the concurrence of assistance, thereby proving an earlier conjecture on concurrence of assistance in networks~\cite{Gou05}.

By showing the upper bound for sharing entanglement is the lower bound for assisting entanglement, this result supports the notion of concurrence of assistance being dual to concurrence, and this lower bound is thus dual to monogamy of entanglement. In proving this result, we have introduced the linear mutual entropy. Although this quantity differs dramatically from mutual entropy, especially in that the linear mutual entropy of product states is nonzero, formally the expression for linear mutual entropy is similar to mutual entropy using the von Neumann entropy. Linear mutual entropy is thus important for studying concurrence and concurrence of assistance and begs further study. Here we have 
obtained tight lower and upper bounds on the sum of pairwise linear mutual entropy with respect to a specific system. These bounds in addition provides interesting entanglement relations between total entanglement across various bipartitions. 

Some implications of dual relations for monogamy of entanglement are bounds on the total entanglement for bipartitions of pure $N$-qubit states with two qubits on one side and the rest on the other in terms of bipartitions with just one qubit on one side of the cut and the rest on the other. These results follow straightforwardly from our theorem and point to hierarchical ways to quantify entanglement in a multi-qubit system, which may elucidate some of the problems with multi-partite quantum networks.
 
\section*{Acknowledgments}
GG appreciates valuable discussions with David Meyer, Peter Stevenhagen, and Nolan Wallach and acknowledges financial support by the National Science Foundation
under Grant No.\ ECS-0202087. SB and BCS acknowledge financial support from 
Alberta's Informatics Circle of Research Excellence (iCORE), 
the Canadian Institute for Advanced Research, the Canadian Network
of Centres of Excellence for the Mathematics of Information Technology and 
Complex Systems (MITACS), the Natural Sciences and Engineering
Research Council, General Dynamics Canada, and the Australian Research Council.

\end{document}